# Self-Frequency Shift of Cavity Soliton in Kerr Frequency Comb

Lin Zhang,[1] Qiang Lin,[2] Lionel C. Kimerling,[1] and Jurgen Michel[1]

[1]*Microphotonics Center and Department of Materials Science and Engineering, Massachusetts Institute of Technology, Cambridge, MA 02139, USA*

[2]*Department of Electrical and Computer Engineering and Institute of Optics, University of Rochester, Rochester, NY 14627, USA*

Emails: linzhang@mit.edu, qiang.lin@rochester.edu

**Abstract:** We show that the ultrashort cavity soliton in octave-spanning Kerr frequency comb generation exhibits striking self-adaptiveness and robustness to external perturbations, resulting in a novel frequency shifting/cancellation mechanism and gigantic dispersive wave generation in response to the strong frequency dependence of Kerr nonlinearity, Raman scattering, chromatic dispersion, and cavity Q. These observations open up a great avenue towards versatile manipulation of nonlinear soliton dynamics, flexible spectrum engineering of mode-locked Kerr frequency combs, and highly efficient frequency translation of optical waves.

Optical frequency combs generated in monolithic devices due to Kerr nonlinearities, also called Kerr frequency combs, have attracted significant interest recently [1-10]. Broadband frequency combs have been produced in high-Q microresonators made with a variety of material platforms and device geometries [2-10], showing great potential for broad applications, e.g., frequency metrology [11], precision spectroscopy [12], optical communications [13], and signal processing [7]. In spite of these technical developments, the underlying mechanism, however, turns out to be fairly complicated, which has attracted intensive investigations in the past few years [14-22]. Very recent work indicates that the formation of cavity soliton is responsible for the mode locking of generated frequency combs [17-22].

Optical solitons represent a fascinating manifestation of nonlinear optical phenomena in nature [23]. Its evolution in optical fibers and integrated nanowire waveguides results in the interesting supercontinuum generation [24, 25], rogue waves [26], gravity-like effect [27], frequency-shift cancellation [28], frequency blue shift [29-31], among others.

However, in the context of optical cavities (e.g., mode-locked lasers), the regenerative self-adaptive nature of the systems leads to very distinctive soliton behaviors, manifesting themselves as various temporal and spatial dissipative solitons [32]. In particular, high-Q microresonators enable greatly tailorable dispersion [33, 34]. This feature, in combination with the highly nonlinear nature of the system, offers a new testbed to explore the intriguing nonlinear optical dynamics in the regime inaccessible to other devices. Here we show that cavity solitons during Kerr comb generation exhibit remarkable self-adaptive characteristics with unique device dispersion, resulting in novel frequency shifting and cancellation mechanisms and radiation emission characteristics, which have never been observed before in other nonlinear optical systems.

As a result of balanced interaction between the group-velocity dispersion and self-phase modulation, optical soliton exist in both waveguides and resonators. Stable soliton propagation can be disturbed by loss, higher-order dispersion, and other nonlinear effects such as stimulated Raman scattering (SRS) and Kerr self-steepening (KSS). In a waveguide, a majority of these detrimental effects lead to soliton breakup, while SRS results in a self-frequency shift of soliton [23]. However, in Kerr frequency comb, we find that the formed cavity soliton is able to maintain its entity and integrity, by adjusting its spectral location.

We choose a specific nonlinear cavity to describe the underlying physics, whose schematic is shown in Fig. 1a. The microresonator exhibits a flat and low dispersion over an octave-spanning bandwidth, and two zero-dispersion frequencies (ZDFs) are located near 130 and 263 THz, respectively (Fig.1b). Finite-element simulation shows that the cavity has very small radiation loss over a 200-THz spectral range (Fig.1c). However, its external coupling to the straight waveguide is much more dispersive. (This effect is neglected in all previous investigations of Kerr combs, although having a striking impact on the soliton dynamics as shown below.) Consequently, the device is over coupled at low frequencies but becomes strongly under coupled at high frequencies, with a loaded cavity Q changing by more than two orders of magnitude from below $10^4$ to above $10^6$ as frequency increases from 120 to 320 THz. Moreover, over such a broad spectral range, the Kerr nonlinear coefficient is also strongly frequency dependent (Fig. 1b), varying by nearly one order of magnitude. All these frequency dependences would have a significant impact on ultrashort soliton dynamics.

To explore the soliton dynamics, we simulate comb generation using the driven and damped nonlinear Schrödinger equation [15, 35] (also called the Lugiato-Lefever equation [17, 19]) as follows:

$$t_R \frac{\partial E}{\partial t} = \sqrt{\kappa_0} E_{in} + l[K(E) + R(E)] - \left( \frac{\alpha}{2} + \frac{\kappa}{2} - j\delta_0 + jl \sum_{m=2}^{\infty} \frac{(-j)^m \beta_m}{m!} \frac{\partial^m}{\partial \tau^m} \right) E$$

where $E=E(\tau,t)$ describes the intra-cavity optical field, $\tau$ and $t$ are the fast and slow times, and $t_R$ is the round-trip time. $E_{in}$ represents the input pump field at frequency $\omega_0$ launched into a cavity mode at a resonance frequency of $\omega_n$ (pump power $P_{in} = |E_{in}|^2$), with a coupling coefficient of $\kappa_0$, and a pump-cavity phase detuning of $\delta_0 = \tau_0 \cdot (\omega_n - \omega_0)$. $\beta_m$ represents the $m$th dispersion coefficient. We include the full frequency dependences of the chromatic dispersion, the power loss per round trip $\alpha$, and the power coupling coefficient κ (Fig. 1b-d). We employ all-order dispersion (AOD) terms in frequency domain as in [17]. Nonlinear Kerr and Raman terms are given as

$$K(E) = -i\gamma_K(1 - i\tau_{shock\_K}\frac{\partial}{\partial\tau})E|E|^2$$

$$R(E) = -i\gamma_R(1 - i\tau_{shock\_R}\frac{\partial}{\partial\tau})[E\int_{-\infty}^{\tau}h_R(\tau-\tau')|E|^2d\tau']$$

where the Kerr nonlinear coefficient $\gamma_K$, the Raman gain coefficient $\gamma_R$, and the Raman response function $h_R$ are defined the same way as in [36]. The Kerr and Raman shock times $\tau_{shock\_K}$ and $\tau_{shock\_R}$ represent the derivatives of $\gamma_K$, and $\gamma_R$, with respect to frequency [24]. In our simulations, the temporal step is 1 fs.

Figure 2 shows an example of cavity soliton dynamics with a pump at 220 THz, where the loaded Q is $5.7\times10^5$, with a finesse of 593, and the nonlinear coefficient γ is 1.2 /(W·m). We specify the pump power by setting the normalized pump X, as defined in [18]. To excite cavity soliton for mode-locked Kerr combs [22], we slightly red-detune the pump every 20 ns into the resonance.

As shown in Fig. 2a, with a normalized pump of X=600, the comb is generated from modulational instability and evolves into a mode-locked regime [22]. Two striking features appear during the comb evolution. First, an extremely strong dispersive wave is generated around 350 THz, with a magnitude as high as the pump inside the cavity (Fig. 2b). Its spectral width is so narrow that it only occupies a few cavity modes. Although the dispersive wave generation (DWG) is initiated by the Cherenkov radiation from a soliton, the emission of such intense dispersive wave is a direct consequence of the high-Q nature of the cavity at high frequencies, which dramatically enhances DWG and stores it efficiently to a significant magnitude. Because of the under coupling to the waveguide, the comb spectrum out of the cavity shows much weaker DWG (Fig. 2b). This points to an important consequence that was neglected in the past: Due to the frequency-dependent external coupling, the output comb spectrum may not accurately reflect the physics inside the resonator, and special attention is required to interpret the experimental measurements. From the application perspective, such highly efficient DWG may find novel applications in large-scale frequency translation of continuous waves, while mediated by soliton dynamics.

Second, the major portion of the spectrum gradually shifts from being centered at the pump frequency during modulational instability to lower frequencies as a cavity soliton is formed. Interestingly, even with the presence of the extremely strong dispersive wave, a stable soliton is still formed, with a full-width at half maximum of 10.6 fs and a pulse peak power of 377 W (Fig. 2c). However, the soliton spectrum is shifted away from the pump by 16 THz (i.e., 107 nm) in Fig. 2b. At a first glance, one might attribute the frequency shift to the frequency recoil [25] from the strong dispersive wave. However, we find that there is much more profound physics underlying such cavity soliton frequency shift, which turns out to be a combined effect of five different physical mechanisms.

First, with a wideband flat dispersion profile, as shown in Fig. 1a, the cavity soliton is able to maintain its integrity, by moving its spectrum towards the center of anomalous low-dispersion band, i.e., a frequency band that better accommodates its spectrum. Such dispersion-induced soliton frequency shift can be along either spectral direction, depending on the pump frequency. This is in contrast to the case in an optical waveguide, where a soliton may break due to higher-order dispersions, as shown for a waveguide with a similar dispersion [37].

Second, the frequency-dependent Q-factor (FDQ) blue-shifts the soliton spectrum towards where the Q is higher. Although the four-wave mixing process underlying the comb generation depends sensitively on the cavity Q, FDQ does not break up the soliton which, instead, maintains a symmetric spectrum with its spectral peak shifted. Note that the spectrum output from the cavity can be quite asymmetric due to the frequency-dependent external coupling. This reminds us again about the importance of correct interpretation of the cavity output, since a FDQ is common to a waveguide-coupled integrated resonator.

Third, the frequency-dependent nonlinear coefficient, manifested as KSS, introduces a blue shift of a cavity soliton spectrum. In a straight waveguide, KSS distorts both soliton waveform and spectrum via intensity-dependent group velocity, and a soliton cannot be supported in the presence of strong KSS [37]. In contrast, a cavity soliton has its waveform "reshaped" via interference with the incoming pump every time when it passes the coupling area between a cavity and a waveguide. The soliton waveform can be maintained, corresponding to a symmetric spectrum as illustrated in Fig. 2b. KSS is then translated into a blue shift of the comb spectrum.

Fourth, DWG produces a frequency shift of Kerr combs via a spectrum recoiling effect [25]. In a waveguide, DWG-induced frequency shift can be bi-directional, depending on the relative spectral location of the soliton with respect to the ZDF. However, in a cavity, due to FDQ, the dispersive wave at high frequencies is much stronger than that at low frequencies. Consequently, the DWG-induced frequency shift is dominated by a red shift. One unique feature of a high-Q microcavity is that the dispersive wave recycles and accumulates inside the cavity for a long time, leading to repeated interaction with a soliton, in contrast to a single-pass waveguide where a soliton moves away from the dispersive wave due to group velocity mismatch [24]. Thus, DWG in a cavity produces stronger frequency recoil than in a single-pass waveguide.

Finally, SRS introduces a red shift of the soliton spectrum via intra-pulse Raman amplification. Here, we find that SRS also helps stabilize the cavity soliton formation, as discussed below.

The overall soliton spectrum is the result of competition among these five physical mechanisms. To isolate the influence of each mechanism underlying cavity soliton frequency shift, we individually turn off the effects in our simulations and compare how the soliton spectrum varies as the pump power is increased. Two pump frequencies at 180 and 220 THz are employed, which are almost symmetric around the center of the low-dispersion band in Fig. 1b. Due to the dramatic FDQ, the pump power needs to be varied over frequency. Therefore, the comb frequency shift cannot be compared as a function of pump power. Instead, a new figure of merit for soliton-based mode-locked Kerr combs is defined as the product of soliton peak power and 3-dB bandwidth, which we call power-bandwidth product (PBP). The PBP value of a comb measures both comb power and its bandwidth with the determined comb flatness from a soliton spectrum.

Figure 3a shows the comb frequency shift (defined as the separation between the soliton center frequency and the pump frequency), when we pump at 180 THz. In general, the frequency comb is fairly centered at the pump, with a small frequency shift of 2~4 THz for a broad range of pumping condition. However, detailed analysis shows that this is a balanced cancellation of soliton frequency shifts resulting from various mechanisms. As shown in Fig. 3a, dispersion induces a strong blue shift of soliton frequency by about 10~20 THz, which is further dramatically increased to 25~40 THz by FDQ. However, it is considerably compensated by the red shift from both the intrapulse SRS and the frequency recoil from the high-frequency dispersive wave, resulting in a net blue shift of about 8~15 THz. Surprisingly, KSS does not further blue shift the soliton spectrum. Instead, it red-shifts the soliton frequency, leading to a small overall frequency shift of only 2~4 THz. This is because KSS corresponds to a frequency-dependent nonlinear gain, leading to a greatly enhanced dispersive wave at high frequency. The intense dispersive wave in turn produces strong frequency recoil which not only compensates the KSS-induced blue shift on the soliton itself, but also further red-shifts the soliton spectrum to compensate the blue shift from other mechanisms. It is clear from this example that the cavity soliton frequency shift in Kerr comb generation is a well-balanced result of various competing mechanisms.

The competing frequency shifting/cancellation can be seen in Fig. 3b as well, when pumping at 220 THz. All mechanisms except FDQ introduce a red shift of the cavity soliton. Cavity dispersion now shifts the soliton frequency towards red by about -10 THz. However, the FDQ-induced blue shift is so strong that it compensates the dispersion-induced red shift, resulting in a net blue shift of 10~17 THz. The soliton center frequency is moved to 237 THz at PBP = 158 W·PHz, close to the ZDF at 263 THz. FDQ not only affects the spectrum output (Fig. 2c), but is also the only physical mechanism responsible for a soliton blue shift. The effect of DWG becomes much stronger compared with the case in Fig. 3a, since the soliton spectrum is closer to the high-frequency dispersive wave. Therefore, the overall soliton frequency is shifted towards red by 10~20 THz for a PBP value of 10~40 W·PHz. SRS not only red shifts the soliton frequency, but it also benefits the soliton formation and comb mode locking. That is, the high-frequency DWG has significant impact on the soliton due to the FDQ and KSS as well as the regenerative nature of cavity solitons. However, SRS moves the soliton spectrum away from the high-frequency dispersive wave, greatly reducing its impact. Consequently, SRS functions as a balancing mechanism to help stabilize soliton formation and comb generation. Detailed simulations show that, without SRS, the DWG would be so strong that the soliton becomes unstable.

AOD plays a significant and unique role in determining the shift of a cavity soliton in a Kerr comb. In Fig. 3, the AOD-induced frequency shift is always towards the center of the anomalous dispersion band, where the whole spectrum of a few-cycle cavity soliton is better accommodated. To more intuitively illustrate the role of AOD, we conduct two investigations, as shown in Fig. 4.

First, we fix the pump frequency at 180 THz and increase the normalized pump X from 100 to 700. Figure 4a and 4b show that the soliton spectrum quickly moves to a higher frequency as the pump increases, but its center saturates at 202 THz, i.e., the center of the anomalous dispersion band, when X becomes larger than 300, resulting in a soliton frequency shift of 22 THz. It is interesting to note that the soliton is able to maintain its integrity even with such a large value of the normalized pump, clearly showing the resilience of the soliton. In practice, however, this does not necessarily infer a high pump power, since X is scaled by the nonlinear coefficient. For resonators made with highly nonlinear materials (e.g., silicon or chalcogenides, with a nonlinear coefficient one or two orders higher than $Si_3N_4$), the real value of the pump power is expected to be below 0.5 W.

Second, we fix the normalized pump X=600 and scan the pump frequency from 150 to 235 THz. As shown in Fig. 4c and 4d, despite such large sweeping of the pump frequency, the center of the comb spectrum remains fairly intact. Consequently, the spectrum center of the comb can be separated quite far from the pump, with a maximum value of 31.5 THz, when the pump is located at 160 THz, corresponding to a 308-nm shift in wavelength.

In conclusion, we have shown that the cavity soliton in octave-spanning Kerr frequency comb generation exhibits intriguing dynamics and spectral characteristics, as a result of a competition of frequency shifts originating from different physical mechanisms. Usually, the spectral location of a Kerr comb is deemed to be a result of pump placement because of the underlying four-wave mixing process. However, we showed that in mode-locked octave-spanning Kerr combs, various perturbation mechanisms play critical roles in the dynamics of a cavity soliton. After formed, the soliton becomes an integrated entity fairly independent from the pump, exhibiting remarkable self-adaptiveness and resilience. These observations not only reveal rich soliton dynamics, but they may also offer a unique avenue for spectral engineering of Kerr frequency comb that would find broad practical applications. For example, one can flexibly moves a Kerr comb to a desired spectral band, where the comb might be difficult to produce otherwise, by tailoring chromatic dispersion, engineering waveguide-to-cavity coupling, controlling the strength of the dispersive wave with a certain spectrally dependent loss and/or suppressing intra-pulse Raman scattering in a crystalline material (e.g., silicon). This is particularly relevant to comb engineering in the spectral regions (e.g., the mid infrared), where the availability of an appropriate pump laser is quite limited.

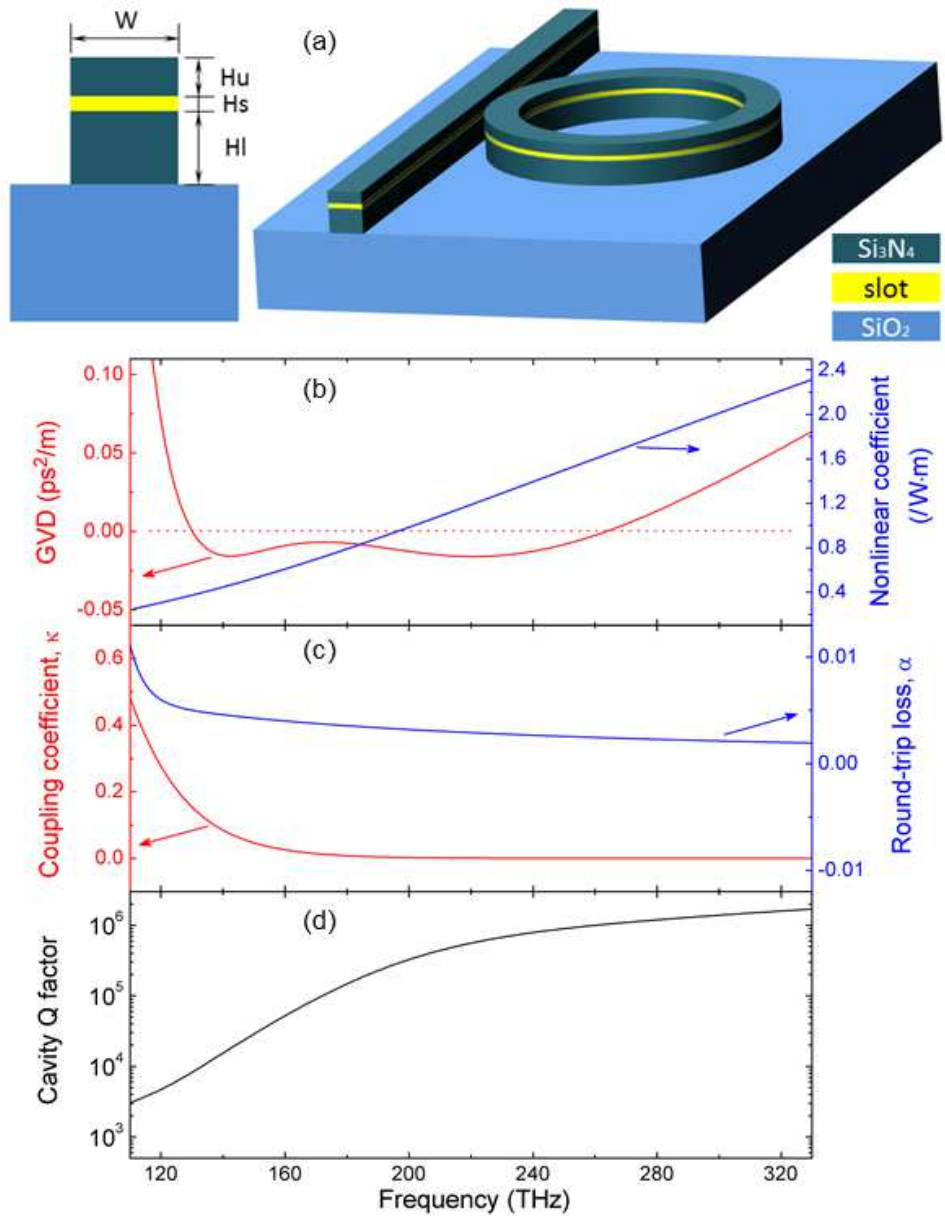

Figure 1 (a) A microresonator has two $Si_3N_4$ layers ($Hu$ = 920 nm and $Hl$ = 480 nm) and a $SiO_2$ layer ($Hs$ = 156 nm) for dispersion engineering, on 3-µm-thick buried oxide. The resonator has a waveguide width of W = 1.3 µm and a radius of 114 µm (corresponding to a free spectral range of 200 GHz and a round-trip time of 5 ps). It is coupled to a waveguide with the same structure separated by 450 nm, which produces critical coupling at ~200 THz. The frequency dependences of its dispersion, nonlinearity, coupling coefficient, round-trip loss and loaded Q-factor are shown in (b), (c), and (d).

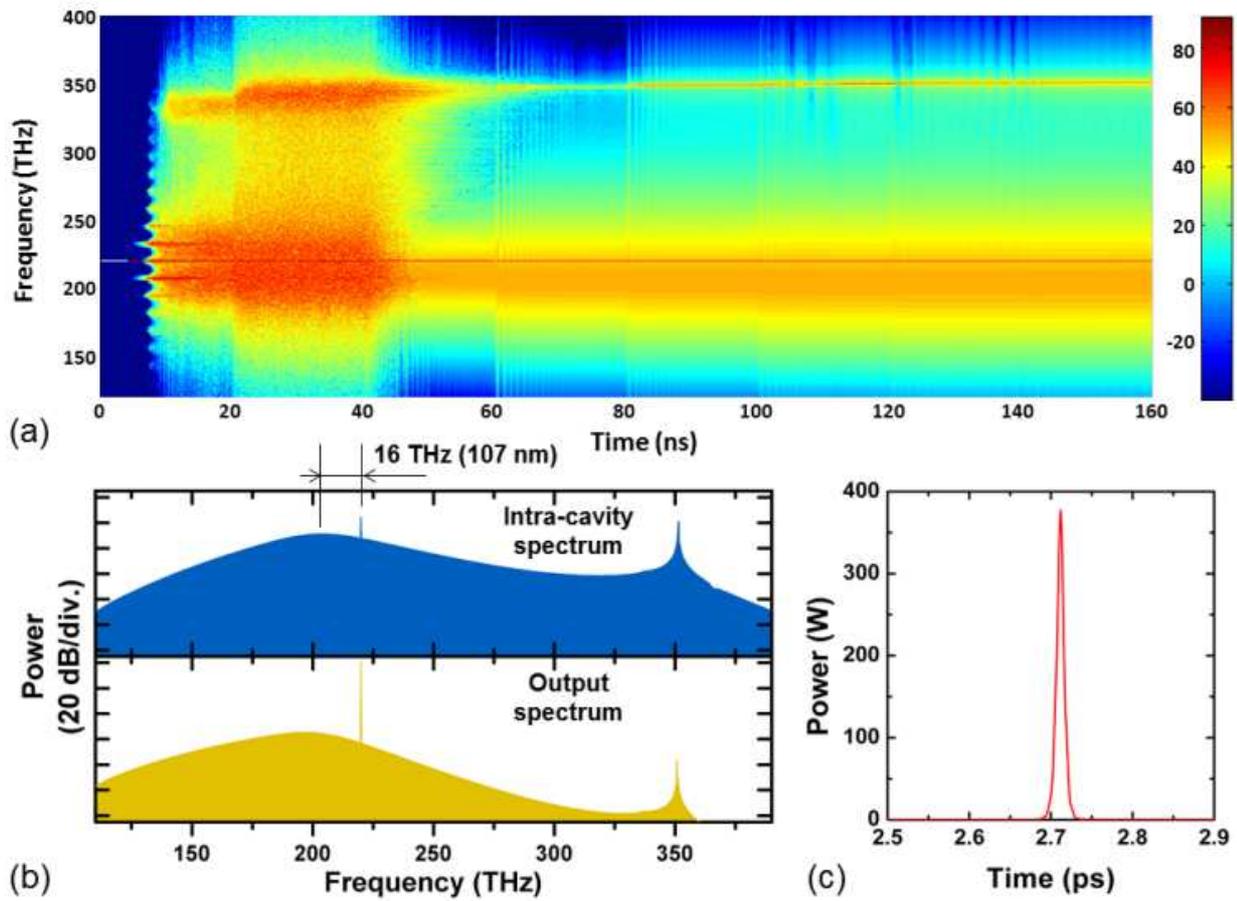

Figure 2 (a) Comb spectrum evolution as pump frequency is slightly detuned into resonance with a step of 20 ns. The comb is red-shifted away from the pump, and a dispersive wave as strong as the pump is generated near 350 THz. At 160 ns, intra-cavity and output comb spectra are shown in (b) and a 10-fs cavity soliton is obtained (c).

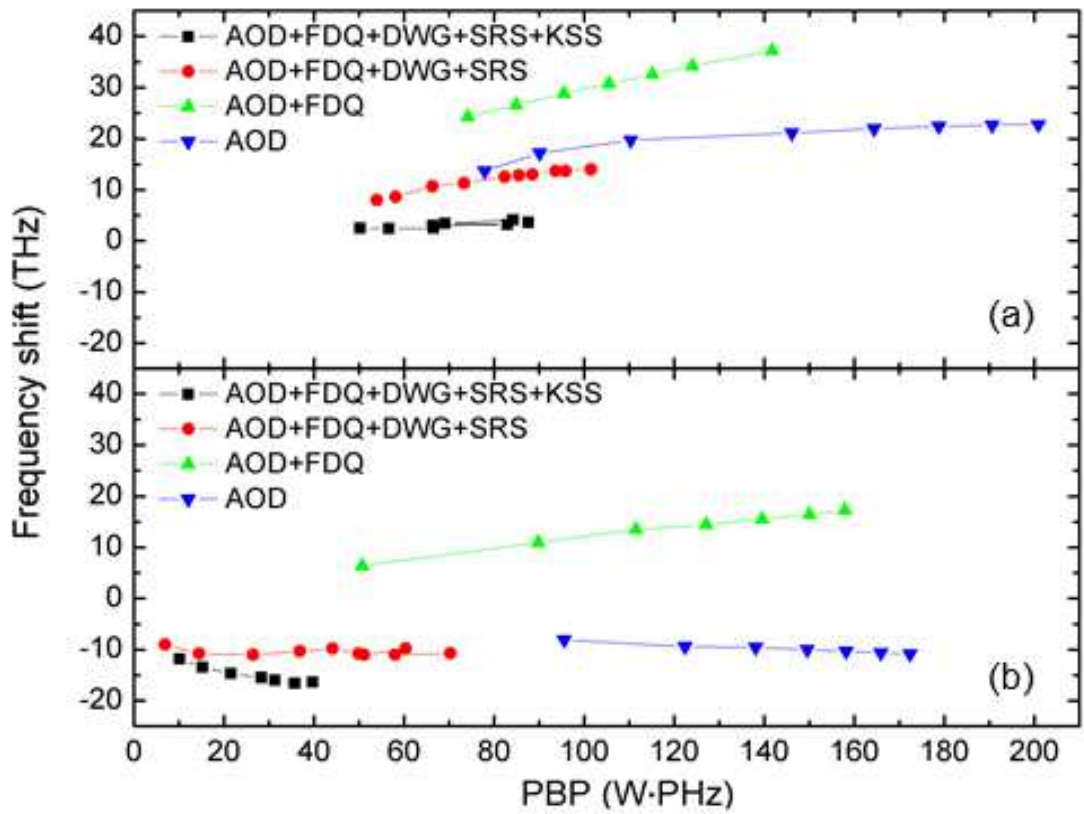

Figure 3 Frequency shift of a cavity soliton in Kerr combs with different soliton perturbation effects and a pump placed at (a) 180 THz and (b) 200 THz, respectively.

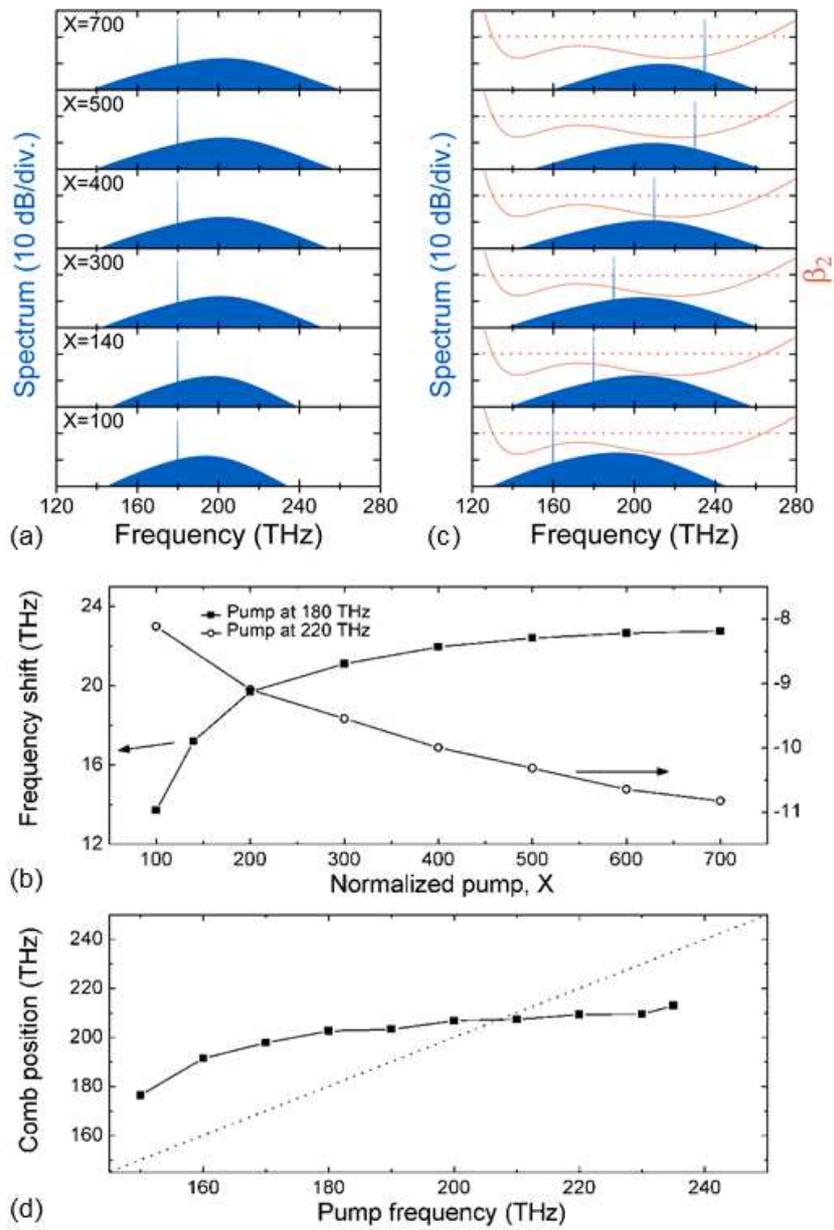

Figure 4 (a) Comb spectrum evolution with different normalized pump X values at 180 THz and according frequency shift versus X in (b). (c) Comb spectrum evolution with scanned pump frequency for X = 600 and according comb position versus pump frequency in (d).